\documentclass[%
reprint,
superscriptaddress,
groupedaddress,
frontmatterverbose, 
nofootinbib,
 amsmath,amssymb,
 aps,
]{revtex4-1}

\usepackage{graphicx}% Include figure files
\usepackage{dcolumn}% Align table columns on decimal point
\usepackage{bm}% bold math
\usepackage{hyperref}% add hypertext capabilities
\usepackage[mathlines]{lineno}% Enable numbering of text and display math
\begin{document}

%\preprint{APS/123-QED}

\title{Cosmic censorship conjecture in some matching\\ spherical collapsing metrics}% Force line breaks with \\


\author{Ramon Lapiedra}
 \email{ramon.lapiedra@uv.es}
\affiliation{%
Departament d'Astronomia i Astrof\'{\i}sica, Universitat de
Val\`encia, 46100 Burjassot, Val\`encia, Spain
}%
\affiliation{%
Observatori Astron\`omic, Universitat de
Val\`encia, E-46980 Paterna, Val\`encia, Spain.
}%
\author{Juan Antonio Morales-Lladosa}%
 \email{antonio.morales@uv.es}
\affiliation{%
Departament d'Astronomia i Astrof\'{\i}sica, Universitat de
Val\`encia, 46100 Burjassot, Val\`encia, Spain
}%
\affiliation{%
Observatori Astron\`omic, Universitat de
Val\`encia, E-46980 Paterna, Val\`encia, Spain.
}%

\date{\today}% It is always \today, today,

\begin{abstract} A physically plausible Lema{\^{\i}}tre-Tolman-Bondi  collapse in the marginally bound case is considered. By ``physically plausible'' we mean that the corresponding  metric is ${\cal C}^1$ matched at the collapsing star surface and further that its {\em intrinsic} energy is, as due, stationary and finite. It is proved for this Lema{\^{\i}}tre-Tolman-Bondi  collapse, for some parameter values,  that its intrinsic central singularity is globally naked, thus violating the cosmic censorship conjecture with, for each direction, one photon, or perhaps a pencil of photons, leaving the singularity and reaching the null infinity. Our result is discussed in relation to some other cases in the current literature on the subject in which some of the central singularities are globally naked too.

\end{abstract}


\maketitle

%\tableofcontents

%%%%%%%%%%%%%
%           %
%  INTRODUCTION %
%           %
%%%%%%%%%%%%%

\section{Introduction }
\label{sec:intro}
Spherical inhomogeneous dust collapse has been extensively studied in the past, paying special attention to the final stages of the evolutionary process. Behind these studies there usually  exists an extra motivation: to confront the validity of the Penrose conjecture \cite{Penrose-79} (censoring the nakedness of  essential space-time singularities) with the singularities developed in specific collapsing situations. For a select set of pioneering work in this context, see, for instance,  Refs. \cite{Eardley-Smarr-1979,Christodoulou-84,Hellaby-Lake-1988,Joshi-Dwivedi-93,Singh-Joshi-96}, which paved the way to delimitate the hypothesis which ensures the validity of the aforementioned Penrose conjecture.

Recently, some spherically symmetric collapsing metrics have been considered in Ref. \cite{Jhin-Kau-2014} (see also the work in  
Refs. \cite{Eardley-Smarr-1979} and \cite{Singh-Joshi-96}),  in order to show that some of their central singularities can violate the Penrose conjecture \cite{Penrose-79,Eardley-Smarr-1979}. In other words, these central singularities could be, against the Penrose conjecture, global naked singularities, i.e., they could be seen from null future infinity. See, for instance ref. \cite{Singh-Joshi-96}, section 4, for a distinction between {\em local} and {\em global} naked singularities. Here, we are only concerned by the possible existence of global naked singularities.

For some authors (see, for instance, Ref. \cite{Singh-99}), to elucidate whether there are in nature naked singularities or not is important since in the affirmative case their vision could give us some clues about how to change the theory of General Relativity in order to avoid these singularities, leading to a, quantum or not, modification of the theory.

In the present paper, we consider the marginally bound case of the dust Lema{\^{\i}}tre-Tolman-Bondi (LTB) family of Einstein equations solutions \cite{Lemaitre-1933,Tolman-1934,Bondi-1947} (see also Refs. \cite{PlebKra,Exact-2003}). We will chose a subfamily made of the particular solutions satisfying the Lichnerowicz matching conditions \cite{Lichnerowicz} with the exterior Schwarzschild metric at the collapsing star%  	
%
%%%%FOOTNOTE-1
%
\footnote{Throughout the paper the term ``star'' refers to any uncharged, spherical, non rotating, finite mass cloud.\label{star}}
%
%%%%END- FOOTNOTE-1
%
%
surface. That is,  in an admissible%
%
%%%FOOTNOTE-2
%
\footnote{Following Ref. \cite{Lichnerowicz}, the term ``admissible'' designates a coordinate system of a ${\cal C}^2$ class (atlas) manifold structure  describing the space-time.\label{admissible}}
%
%%%%END- FOOTNOTE-2
%
coordinate system, the metric is assumed to be class ${\cal C}^1$; i. e., the metric and its first derivatives are assumed to be continuous across this surface.
Perhaps this condition is not always a physically realistic one, but in our opinion, it could be worth to exploring its consequences, 
as we do in the present case.

Furthermore, we impose the physical condition that these metrics have a finite stationary {\em intrinsic}  energy (see Appendix \ref{ap-A}), and finally for the sake of simplicity we choose a simple metric of this particular subfamily of metrics. Hereafter, we name this chosen metric the $\xi$-metric by reasons that will appear later, when we introduce the $\xi$ parameter in  Sec. \ref{sec:4}.

Our main result is that for this $\xi$-metric, and for some parameter values,  the intrinsic central singularity is a globally naked singularity; that is, given a 3-space direction, one outgoing radial null geodesic (or perhaps a pencil of such geodesics) leaves this singularity and reaches the future null infinity. On the other hand, in Ref. \cite{Singh-Joshi-96}, the authors raise the following question: ``Could it be that the initial distributions which lead to naked singularities are not astrophysically realizable?'' Thus, our result suggests that such distributions are astrophysically realizable. Some previous results in Refs. \cite{Eardley-Smarr-1979}, \cite{Singh-Joshi-96} and \cite{Jhin-Kau-2014}, for some marginally bound LTB metrics, seem to support the same suggestion, although differently to our case all but one of these metrics leading to the previous results do not fulfill all the ${\cal C}^1$ matching requirements across the star boundary. Then, we could confirm that the cosmic censorship conjecture would become violated.

This is the paper's outline. In Sec. \ref{sec:2} we obtain the $\xi$-metric, a LTB marginally bound solution obeying the ${\cal C}^1$ matching conditions with vanishing {\em intrinsic} energy.  Section {\ref{sec:3} revisits a sufficient condition for the global nakedness of the central singularity. In Sec. \ref{sec:4},  we prove that this $\xi$-metric fulfills, for some parameter values, this sufficient condition and, in Sec. \ref{sec:5}, we analyze numerically this global nakedness with the help of Mathematica.  The last section, Sec. \ref{sec:6}, is devoted to final considerations. Detailed calculations concerning the {\em intrinsic} energy of the $\xi$-metric have been included in Appendix \ref{ap-A}. The causal character of the apparent horizon of this metric is analyzed in Appendix \ref{ap-B}.

We take $G=c=1$ for the gravitational and the speed of light constants.

%%%%%%%%%%%%%
%           %
%  SECTION  2  %
%           %
%%%%%%%%%%%%%

\section{matching the LTB marginally bound collapse}
\label{sec:2}

As it is well known, when referred to Gauss coordinates adapted to the spherical symmetry, in the marginally bound case the metric element of the dust LTB metrics can be written \cite{PlebKra,Exact-2003} (signature $+ 2$)
\begin{equation}\label{LTB0}
ds^2 =-d\tau^2 + A'^2 d\rho^2 +A^2 (d\theta^2 + \sin^2 {\hspace{-0.5mm}}\theta \, d\phi^2),  
\end{equation}
with $A=A(\tau, \rho)$ and $A' \equiv \partial_\rho A$. The general expression for $A$, the solution of the Einstein field equations,  is
\begin{equation} \label{k0}
A(\tau, \rho) = \Big(\frac{9}{2} M\Big)^{1/3} (\tau - \psi)^{2/3}, 
\end{equation}
where $M=M(\rho)$ and $\psi=\psi(\rho)$ are two arbitrary functions of $\rho$, $M(\rho)$ representing the enclosed partial mass in the sphere of radius $\rho$ and $\psi(\rho)$ representing  the singular time $\tau$  for the $\rho$ shell. The regular coordinate ranges are  $- \infty < \tau < \psi(\rho)$, $0 \leq \rho < \infty$,  
$0 < \theta < \pi$,  and $0 \leq \phi \leq 2\pi$. The 2-surface $\tau = \rho = 0$ and variable $\theta$ and $\phi$ will be referred to as the central singularity.%

We can supplement Eq. (\ref{k0}) with the particular Einstein field equation
\begin{equation} \label{mu}
4 \pi \mu (\tau, \rho) = \frac{M'}{A^2 A'}, \quad M' \equiv \frac{dM}{d\rho}, 
\end{equation}
relating the energy density source, $\mu$, to the metric.

Let us take the commonly used scale $A(0,\rho) = \rho$ (see, for instance, Refs. \cite{Christodoulou-84}; \cite{Joshi-llibre}, p. 245; and \cite{Joshi-Malafarina-2011}, p. 17). This leads to 
\begin{equation} \label{psi}
\psi = \frac{2}{3} \frac{\rho^{3/2}}{\sqrt{2M}}.
\end{equation}
In this gauge, the ${\cal C}^1$ matching conditions with the exterior Schwarzschild metric [see next Eq. (\ref{SchApsi})] 
through the star surface, say $\rho = \lambda$, $M(\rho \geq \lambda) = m = const.$, are in the usual Hadamard notation \cite{Hadamard}, 
\begin{equation}\label{Lichne}
[M]=  [M'] =  [M''] = 0.
\end{equation}
For a detailed proof of this result, see Ref. \cite{LaMo-arXiv-2016}. Notice that in Eq. (\ref{LTB0}), besides $A$, its first derivative $A'$ appears. As a result, the ${\cal C}^1$ 
matching conditions involve the second derivative $A''$ too, which leads finally to the last condition of (\ref{Lichne}), i. e., $[M''] = 0$.

A simple solution of the Eq. (\ref{Lichne}) is%  	
%
%%%%FOOTNOTE-3
%
\footnote{Solution (\ref{Mrho-simple}) is a specially simple case inside a large family of LTB metrics satisfying  Eq. (\ref{Lichne}). See next Eq. (\ref{Mrho2}).  \label{simple}}
%
%%%%END- FOOTNOTE-3
%
\begin{equation} \label{Mrho-simple} 
M(\rho) = 
\begin{cases} 
\displaystyle{m -  m \Big(1- \frac{\rho^2}{\lambda^2}\Big)^3, \quad \rho \leq  \lambda} \\ 
m, \quad \rho \geq \lambda 
\end{cases}
\end{equation}
With this solution we will build, through (\ref{LTB0})--(\ref{psi}), what we have called in the Introduction the $\xi$-metric. Further, this metric has, as due, a stationary and finite {\em intrinsic} energy, as shown in Appendix \ref{ap-A}. Notice that the  $\xi$-metric source is a spherical finite mass, regularly distributed before the eventual collapse. Further, this physical system neither expels nor accretes any mass and neither radiates electromagnetically nor gravitationally. Then, any meaningful kind of energy we can ascribe to it has to be actually stationary and finite as we have demanded, irrespective of how much we approach the physical singularity.

However, before arriving at the basic result of the present section, notice, to begin with, that the expression (\ref{k0}) can be written for the Schwarzschild solution like
\begin{equation}\label{SchApsi}
A = r = \Big(\frac{9m}{2}\Big)^{1/3} (\tau - \psi)^{2/3}, \quad \psi = \frac{2}{3} \frac{\rho^{3/2}}{\sqrt{2m}}, 
\end{equation}
with $r$ the standard static radial coordinate and $m$ the Schwarzschild mass parameter. Then, the generic singularity event, $\tau = \psi(\rho)$, will be visible from the star outside if the leaving photon arrives at the star surface, $\rho = \lambda$, in a time $\tau_\lambda$ of which the corresponding $r$ value given by Eq. (\ref{SchApsi}) is such that 
\begin{equation}\label{r-major2m}
r > 2m.
\end{equation}
This condition gives for $\tau_\lambda$ the inequality
\begin{equation}\label{Sch-taulambda}
\tau_\lambda <  \frac{2}{3} \frac{\lambda^{3/2}}{\sqrt{2m}} - \frac{4m}{3} = \psi(\lambda)- \frac{4m}{3} = \tau_h(\lambda), 
\end{equation}
with $\tau_h(\rho) = \psi(\rho) - \frac{4M}{3}$,  which is called the apparent horizon of the metric (\ref{LTB0}) (see Ref. \cite{Eardley-Smarr-1979}) and is implicitly defined by
\begin{equation}\label{A2M}
A(\tau_h(\rho), \rho) = 2M(\rho).
\end{equation}

In other words, {\em a radial outgoing null geodesic leaving out the generic singular event $\tau = \psi(\rho)$ could only be seen from future null infinity if 
its corresponding photon actually arrives at the star surface and then if its arrival time, $\tau_\lambda$, to this surface satisfies the inequality 
(\ref{Sch-taulambda}), $\tau_\lambda < \tau_h(\lambda)$.}

Nevertheless, for the metric given by (\ref{LTB0}) and (\ref{k0}), there is a well-known result (see, for instance, Ref. \cite{PlebKra} p. 332, at the beginning of Sec. 18.14, and the reference 17 in Ref. \cite{Lake-2015}), according to which if $M' > 0$ all these singularities, out of the central one, are not visible from this future null infinity. That is, all these singularities are dressed ones. Thus, we will concentrate on the possible global nakedness of the remaining singularity, the central one,  of our $\xi$-metric, and then, in the final section, we will compare our result with some well-known results of the present literature on the subject. 
%

%%%%%%%%%%%%%
%           %
%  SECTION  3  %
%           %
%%%%%%%%%%%%%

\section{Sufficient condition for the global nakedness of the central singularity}
\label{sec:3}

In Ref, \cite{Jhin-Kau-2014}, Eq. (26), a sufficient condition for the global visibility of the central singularity,
\begin{equation}\label{consuf}
\frac{\psi'}{M'} > \frac{1}{3}(26 + 15 \sqrt{3}), \quad M'>0,  \quad  \forall \rho\in(0, \lambda), 
\end{equation}
is given for the case of a marginally bound dust LTB metric. Notice that the present notation is different from the one used in Ref. \cite{Jhin-Kau-2014}. 
The justification of the above inequality concerns the behavior of the radial null geodesics across the region 
\begin{equation}\label{A-major2M}
A > 2M, 
\end{equation}
outside the apparent horizon. To make our discussion self-contained,  we give next our version of this justification.

%%%%%%%%%%%%%
%           %
%  Subsection 3A  %
%           %
%%%%%%%%%%%%%

\subsection{Null geodesics from the center}
\label{subsec-3a}
To begin with, the general equation for the radial outgoing null geodesics, $(\tau_g(\rho), \rho)$, for the metric (\ref{LTB0}) is
\begin{equation}\label{geo-radial}
\frac{d\tau_g}{d\rho} = A', 
\end{equation}
where $A'$, having in mind Eq. (\ref{k0}), becomes 
\begin{equation}\label{Aprima}
A' = \frac{1}{3} \frac{M'}{M} \, A + \sqrt{\frac{2M}{A}} \, \psi'.  
\end{equation}
In the region $A > 2M$, let us consider the $k$-lines implicitly defined by the condition
\begin{equation}\label{AkM}
A(\tau_k(\rho), \rho) = k M(\rho), \quad k > 2, 
\end{equation}
which from Eq. (\ref{k0}) is equivalent to
\begin{equation}\label{rhok}
\tau_k(\rho) = \psi(\rho) - \frac{k}{3}\sqrt{2k} \, M(\rho) 
\end{equation}
with $k>2$. The slope of these lines,
\begin{equation}\label{rhok-prima}
\tau_k'(\rho) = \psi'(\rho) - \frac{k}{3}\sqrt{2k} \, M'(\rho),  
\end{equation}
might be compared with the slope of the outgoing radial null geodesics, $\tau_g' = A'$, on the events $(\tau, \rho)$ where both families of lines, $\tau_k(\rho)$ and $\tau_g(\rho)$, intersect. Notice that these intersection events could always exist since they can always be considered the initial condition of a corresponding unique outgoing radial geodesic. Thus, taking  $A=kM$ in Eq. (\ref{Aprima}), we have for these intersecting events:
\begin{equation}\label{rhog-prima}
\tau_g'(\rho)_{|_{kM}} \equiv A'(\tau_k(\rho), \rho) = \sqrt{\frac{2}{k}} \psi'(\rho) + \frac{k}{3} \, M'(\rho),  
\end{equation}

Then,  a  sufficient condition for this geodesic escaping to null  infinity is that for all these intersection events 
of the geodesic lines, $\tau_g$, with some $\tau_{k>2}$ line, with the $\rho$ values belonging to the $(0, \lambda)$ interval, 
we have the following inequality
\begin{equation}\label{compara}
\tau_g'(\rho)_{|_{kM}} < \tau_k' (\rho),  \quad k > 2.
\end{equation}
In fact, from (\ref{rhok-prima}), $\tau'_{k>2} <  \tau'_{k=2} = \tau'_h$, and (\ref{compara}) implies 
\begin{equation}\label{compara2}
\tau'_g (\rho)_{|_{kM}} < \tau'_h (\rho), 
\end{equation}
in this interval. 

Then, in particular, {\em the photon arrives at the star surface at time $\tau_\lambda$, which satisfies 
(\ref{Sch-taulambda}). Consequently,  the photon escapes at the null infinity.}

It remains to prove that such a geodesic starts from the central singularity when Eq. (\ref{compara}) occurs. 
%

%%%%%%%%%%%%%
%           %
%  Subsection 3B  %
%           %
%%%%%%%%%%%%%

\subsection{Null geodesics from the central singularity}
\label{subsec-3b}

From Eqs. (\ref{rhok-prima}) and (\ref{rhog-prima}), the sufficient condition (\ref{compara}) is equivalent to
\begin{equation}\label{compara-k}
\frac{\psi'}{M'} > \frac{k}{3} \, \frac{1+ \sqrt{2k}}{1 - \sqrt{\frac{2}{k}}} \equiv f(k), \quad k>2, 
\end{equation}
which coincides with Eq. (25) in Ref. \cite{Jhin-Kau-2014}, once the corresponding change in notation is taken into account. 

Let us be more precise. Actually, the fulfillment condition (\ref{compara}) for all $\rho \in (0, \lambda)$ implies that the corresponding $k$-line has to be timelike. One can easily arrive at this conclusion by simply drawing the forward outgoing light cone in the assumed intersection event of $\tau_g$ with $\tau_{k>2}$. These timelike lines actually exist because from Appendix \ref{ap-B}, for each $k>2$, the corresponding $k$-line is timelike [as it was implied by (\ref{compara})], provided that Eq. (\ref{compara-k}) be satisfied for all $\rho \in (0, \lambda)$.

For $k>2$, the function $f(k)$ has a  global minimum at $k = k_m = 2 + \sqrt{3}$,  the value of which is
\begin{equation}\label{f(k)-min}
f(k_m)= \frac{1}{3}  (2 + \sqrt{3})^3 = \frac{1}{3} (26 + 15 \sqrt{3})
\end{equation}
according to Eq. (26) in Ref. \cite{Jhin-Kau-2014}. In fact, it is easy to verify that $f''(k_m) >0$.

In particular, the sufficient condition (\ref{compara-k}) will be minimally demanding for $k=k_m$. Then, henceforth, we will put in (\ref{compara-k}) $f(k_m)$, that is we will demand (\ref{consuf}). From this assumption and the above considerations, the following general statement (cf  Ref. \cite{Jhin-Kau-2014} and references quoted therein) can be proved:
{\em For any marginally bound LTB metric (\ref{LTB0}) satisfying  the inequality (\ref{consuf}) with $M$ and $M'$ positive functions in the vicinity of $\rho = 0$ and $M(0) = M'(0) = 0$,  there could exist a pencil of radial null geodesics which come from the central singularity and escape from the star.}

Let us prove this  result step by step: 

(i) The family of lines (\ref{rhok}) intersects the central singularity $\tau_k (0) = \psi(0)$ because  $M(\rho)$ goes to zero when $\rho \to 0$. 

(ii) In the vicinity of $\rho =0$, the slope of every line $\tau_k$ ($k>2$) remains larger than the corresponding slope $\tau'_{g}(\rho)_{|kM}$ for the outgoing radial null geodesic (compare (\ref{rhok-prima}) and (\ref{rhog-prima}) keeping in mind that $M'(0) = 0$). 

(iii) Moreover,  taking into account Eq. (\ref{f(k)-min}), the smoothness of the functions involved in Eq. (\ref{compara-k}) guarantees that an open elementary interval around $k_m$, $(k_m-\epsilon, k_m + \epsilon)$, exists such that Eq. (\ref{compara}) is satisfied, that is,
\begin{equation}\label{interval}
\tau'_g(\rho)_{|_{l M}} <  \tau'_l (\rho) \quad   \forall l \in (k_m-\epsilon, k_m + \epsilon), \, \forall \rho \in(0, \lambda).
\end{equation} 

(iv) Then, let us consider any one of the $k$-lines, $k=l$, and any one of the events on it; let us say the event corresponding  to $\rho = \rho_1$. 
Further, given a direction $\theta, \phi$, consider the virtual unique null outgoing geodesic, say $\tau_g(\rho)_{|_{l}}$, passing through this $\rho_1$ event.  Assume that this virtual geodesic exists actually from $\rho = 0$. Can this geodesic remain over $\tau_l(\rho)$ when $\rho$ goes to zero? No, it cannot, since $(\tau=0,\rho=0)$ 
is the essential central singularity, such that events with $\rho=0$ and $\tau > 0$ are forbidden. Could then  the geodesic run, for $\rho$ 
going to zero, the opposite way, that is, to start from $\rho =0$ below the $l$-line, $\tau_l(\rho)$? No, since in order to arrive at Eq. (\ref{interval}) for $\rho= \rho_1$ we should have, contrarily to Eq. (\ref{interval}), $\tau'_g(\rho)_{|_{l M}} >  \tau'_l (\rho)$ for some $\rho = \rho_2 < \rho_1$. 
But, as remarked above in the present section, referring to Appendix \ref{ap-B}, the $l$-line is timelike. 
Thus, simply drawing the corresponding outgoing light cone for $\rho_2$ one becomes convinced that the last inequality is impossible. In all, the outgoing radial $l$-geodesics, $\tau_g(\rho)_{|_l}$, start from the central essential singularity.

Therefore, a pencil of photons, one photon for each one of the above corresponding $l$ and $\rho_1$ values, would exist and would be emitted from the central singularity and would remain always out of the apparent horizon $A = 2M$ and, consequently, it could be detected outside the star. 

Contrarily, no such a pencil can be present when we consider the light leaving out  the central regular events 
$(\tau <0$, $\rho = 0)$, since given a direction $(\theta, \phi)$ there is a unique radial null geodesic leaving out any regular event. Then, although leaving a door open to the actual existence of that photon pencil leaving out the central singularity, we must admit that such a pencil could be the result of having assumed the actual existence of some virtual photons.

Notice that, in a mathematical terminology, Eq. (\ref{interval}) together with the algebraic conditions $\tau_l(0) = \tau_h(0)$ 
and $\tau_l(\rho) < \tau_h(\rho)$ for all $\rho \in (0,\lambda)$ say that the lines $\tau_l(\rho)$ are subhorizon supersolutions 
of Eq. (\ref{geo-radial}) of which the existence is equivalent to the global naked character of the central singularity (see Ref. \cite{Gi-Gi-Ma-Pi-2003}, Theorem 2.5).  
We have just then proven that $\tau_l(\rho)$ is a set of subhorizon supersolutions of Eq. (\ref{geo-radial}).

%%%%%%%%%%%%%
%           %
%  SECTION  4  %
%           %
%%%%%%%%%%%%%

\section{Proving that the central singularity of the $\xi$-metric is globally naked for some $\xi$ values}
\label{sec:4}

In the present section, we will show, for some parameter values,  that the central singularity $\tau = \psi(\rho = 0)=0$ for the $\xi$-metric (see Sec. \ref{sec:2}) is a global naked singularity, in accordance with a similar result from Ref. \cite{ Christodoulou-84}. Our result will be obtained numerically in the next section, and also  applying the sufficient condition (\ref{consuf}), according to Ref. \cite{Jhin-Kau-2014}, in the present section. 

However, it cannot be obtained from  Ref. \cite{ Christodoulou-84} going to the limiting case where the 3-space curvature vanishes, since this limit does not allow us to recover our $\xi$-metric.

Using inequality (\ref{consuf}), the authors of Ref. \cite{Jhin-Kau-2014} prove the existence of four metrics with a global naked central singularity for four different functions $M(\rho)$,  Eqs. (28), (33), (38), and (43), respectively, of the Sec. V of Ref. \cite{Jhin-Kau-2014}. But these $M$ functions do not fulfill the last condition (\ref{Lichne}), $[M''] = 0$, and then do not fulfill all  the corresponding ${\cal C}^1$ matching conditions across the star boundary, $\rho = \lambda$. Could this non-fulfillness be the reason for the  nakedness and so the reason for the corresponding violation of the cosmic censorship conjecture? The answer is negative, since we are going to see that our $\xi$-metric, which satisfies all conditions (\ref{Lichne}), has a central global naked singularity for large enough  values of the parameter $\xi \equiv \lambda / 2m$.

Let us have in mind Eqs. (\ref{psi}) and (\ref{Mrho-simple}) for $\rho \leq \lambda$. In terms of the dimensionless variable $x=\rho/\lambda \in [0, 1]$,  the mass function and the singularity time lines are given by
\begin{equation}\label{M-normalitzada}
\frac{M(x)}{m} = x^2 (x^4 -3x^2 +3) \equiv x^2 P(x),
\end{equation}
and
\begin{equation}\label{psi-normalitzada}
\frac{\tau(x)}{m} = \frac{\psi(x)}{m} = \frac{4}{3} \xi^{3/2} \sqrt{\frac{x}{P(x)}}, 
\end{equation}
respectively, where $P(x) \equiv x^4-3x^2+3$. Thus, taking into account (\ref{M-normalitzada}) and (\ref{psi-normalitzada}),  the inequality (\ref{compara-k}) becomes:
\begin{equation}\label{consuf-gen}
\xi >   [3f(k)]^{2/3} F(x), 
\end{equation}
where the function of the right hand is
\begin{equation}\label{Fx}
F(x) \equiv  x P(x) \Big(\frac{(1-x^2)^2}{1+ x^2 - x^4}\Big)^{2/3}, 
\end{equation}
which has a maximum value $F_{max} \approx 0.74$ at $x \approx 0.4$ (see Fig. \ref{xi-max}). 
Thus, Eq. (\ref{consuf-gen}) is the expression of the sufficient condition (\ref{compara-k}) for the $\xi$-metric.
In particular, for $k=k_m$,  from Eqs. (\ref{f(k)-min}) and (\ref{consuf-gen}) we obtain
\begin{equation}\label{consuf2}
\xi >   (2 + \sqrt{3})^2 F(x).
\end{equation}

\begin{figure}
\centerline{
\parbox[c]{0.6\textwidth}{\includegraphics[width=0.5\textwidth]{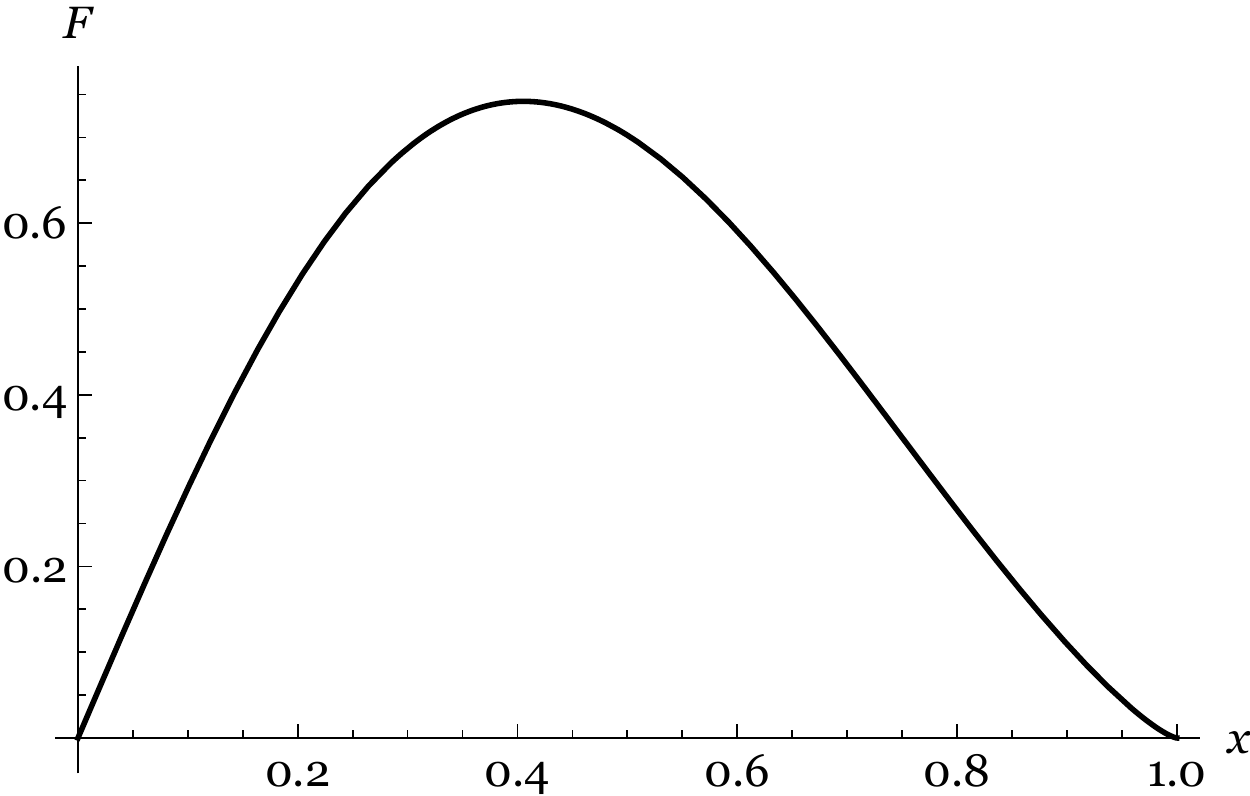}}}
\caption{Plot of $F(x)$ showing its maximum value at $x \approx 0.4$. In the text notation,  $F_{max} \approx F(0.4) \approx 0.74$.
The  central singularity of a $\xi$-metric is globally naked  when the $\xi$ parameter, $\xi = \lambda / 2m$, is larger than $(2 + \sqrt{3})^2 F_{max} \approx 10.33$. 
\label{xi-max}}
\end{figure}
\begin{figure}
\centerline{
\parbox[c]{0.6\textwidth}{\includegraphics[width=0.5\textwidth]{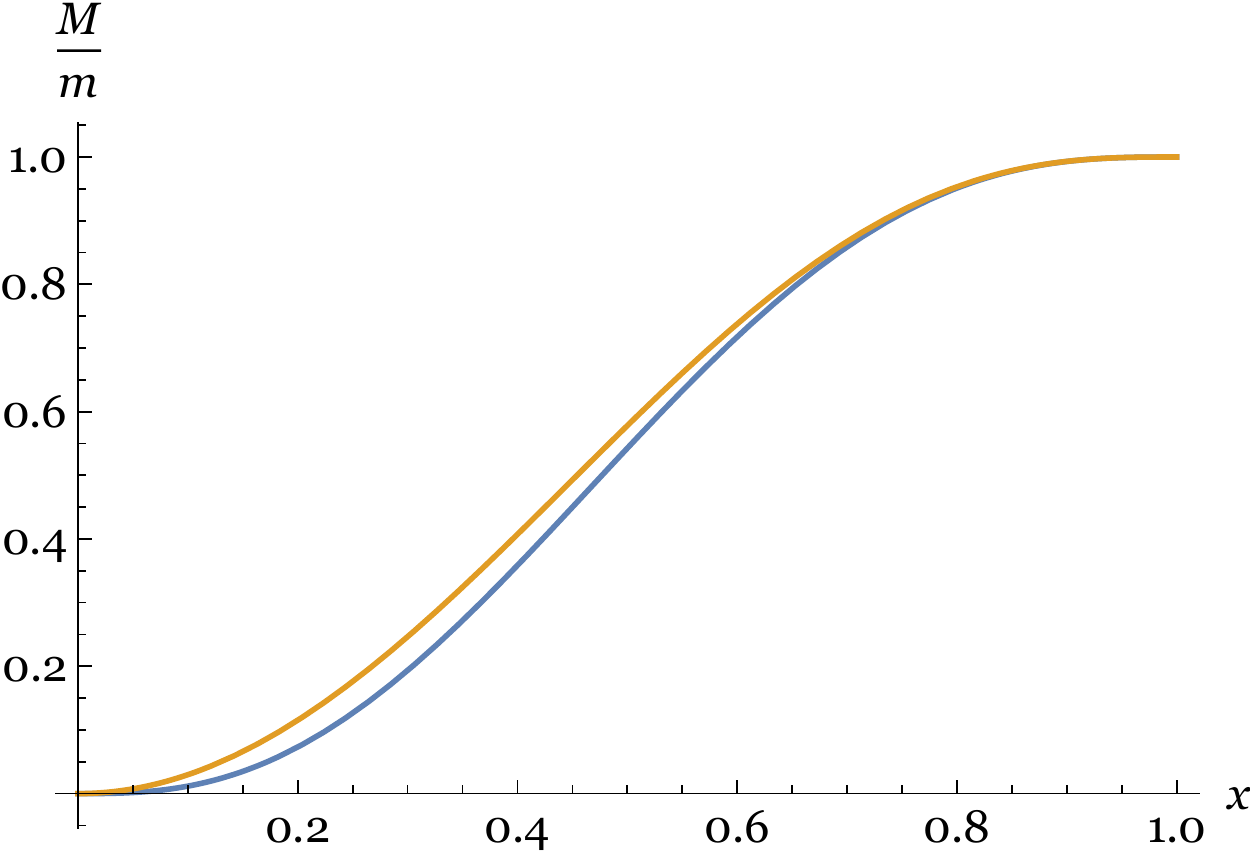}}}
\caption{The upper line (colored in orange) is the normalized mass function of the $\xi$-metric, $\frac{M(x)}{m} = x^2 (x^4 -3x^2 +3)$. The line below that (colored in blue) is the mass function of Eq. (47) in Ref. \cite{Jhin-Kau-2014}, $\frac{M(x)}{m} = x^3 (16 -15 x + 12 x \ln x)$.
\label{dues-M}}.
\end{figure}

Then, for any value of $\xi$ larger than $(2 + \sqrt{3})^2 F_{max} \approx 10.33$,  the corresponding $\xi$-metric has a central global naked singularity.  As discussed at the end of Sec. \ref{sec:3}, this naked singularity leads, for each central direction, to a unique infinite escaping photon or even to a pencil of them.

On the other hand, this value of $F_{max}\approx 0.74$ provides the threshold value of the $\xi$ parameter from which the apparent horizon of the  metric is everywhere spacelike for all $x \in (0,1)$. For a detailed proof of this statement see Appendix \ref{ap-B}. 

The above four $M$ functions of Ref. \cite{Jhin-Kau-2014} can be specified  in terms of local expansions in $\rho$ near the vanishing value of $\rho$: the three first $M$ functions go like $\rho^3$, for small $\rho$ values, and the fourth one goes like $\rho$, while our $M$ function, 
Eq. (\ref{Mrho-simple}), goes like $\rho^2$. 

But in  Ref. \cite{Jhin-Kau-2014}, a fifth case for $M$  is considered (see Eq. (47) in Ref. \cite{Jhin-Kau-2014}), this one leading to another central global naked singularity in the paper and, although it is not mentioned in Ref.  \cite{Jhin-Kau-2014}, the corresponding metric fulfilling the ${\cal C}^1$ matching conditions. Furthermore, its {\em intrinsic} energy is finite and stationary (actually,  it vanishes), as it must be according to the comment at the paragraph that follows Eq. (\ref{Mrho-simple}) in Sec. \ref{sec:2}.  The {\em intrinsic} energy of our $\xi$-metric vanishes too, because in this case, as we have noted, $M \sim \rho^2$, for $\rho \to 0$ (see Appendix \ref{ap-A}). The similar vanishing in the fifth $M$ case of Ref. \cite{Jhin-Kau-2014} comes {\em a fortiori} from the fact that now  $M \sim \rho^3$, for $\rho \to 0$. Actually, the mass function (\ref{M-normalitzada}) of the $\xi$-metric is slightly  greater than the mass given by Eq. (47) in Ref. \cite{Jhin-Kau-2014} (see Fig. \ref{dues-M}).

In all, the Penrose cosmic censorship conjecture becomes violated at least for two plausible --${\cal C}^1$ matched and with a finite, stationary, {\em intrinsic} energy-- metrics belonging to the marginally bound dust LTB family, one of these two metrics having already been proposed in Ref. \cite{Jhin-Kau-2014}, although the authors had not noticed that the proposed metric was a ${\cal C}^1$ matching metric with a finite, stationary, vanishing intrinsic energy.

%%%%%%%%%%%%%
%           %
%  SECTION  5  %
%           %
%%%%%%%%%%%%%

\section{Showing by numerical calculation that the central singularity of the $\xi$-metric is globally naked for some $\xi$ values}
\label{sec:5}
Inequality (\ref{consuf2}) is a sufficient condition for central global nakedness, but not a necessary  condition. Then, helped by Mathematica, we numerically calculate some of the outgoing central null geodesics of our $\xi$-metric for different values of the $\xi$ parameter. We will show the existence of this nakedness for $\xi$ values lower than the above $(2 + \sqrt{3})^2 F_{max} \approx 10.33$ value. 

The outgoing radial null geodesics $(x, y(x))$ of the marginally bound  LTB metric (\ref{LTB0}) are the solution of the ordinary differential equation%  	
%
%%%%FOOTNOTE-4
%
\footnote{To perform numerical integration and graphic representation, normalized variables $(x, y)=(\rho/\lambda, \tau/m)$ are used for convenience. Note the irrelevant, but graphically convenient, order change with respect the starting $(\tau, \rho)$ coordinates. \label{no-shell-crossing}}
%
%%%%END- FOOTNOTE-4
%
%

%
\begin{equation}\label{eq-geo}
 y'(x) = \frac{1}{m}A'(y(x), x), 
\end{equation}
where $y(x) \equiv \tau_g(x)/m$ and now the prime stands for the derivative of $y$ with respect to $x$. Here,  since we are dealing with the specific case of the $\xi$-metric, we must use Eq. (\ref{k0}) with $M$ given by Eq. (\ref{M-normalitzada})  and $\psi$ by Eq. (\ref{psi-normalitzada}), in the interior of the star,%  	
%
%%%%FOOTNOTE-5
%
\footnote{Notice that, inside the star, $\psi$ and $M$ are both increasing functions. Then,  Eq. (\ref{Aprima}) implies that $A'$ is always positive. Consequently shell crossing singularities (see Ref. \cite{PlebKra}, p. 321) will not occur during the collapse. In addition, the proper energy density $\mu \equiv \mu(\tau, \rho)$,  that is, according to Eq. (\ref{mu}), $4 \pi \mu = M'/(A^2 A')$,  is everywhere regular (except for the essential singularity $A=0$). These properties could reinforce the belief in the goodness of the $\xi$-metric. \label{no-shell-crossing}}
%
%%%%END- FOOTNOTE-5
%
%
and then the second member of (\ref{eq-geo}) has the expression
\begin{eqnarray}\label{second-member}\nonumber
\frac{1}{m}A'(y, x)& = & \Big(\frac{2}{P}\Big)^{2/3} \Big[ 3^{2/3} \frac{(1-x^2)^2}{x^{1/3}} \Big(\frac{4}{3} \xi^{3/2}\sqrt{\frac{x}{P}}-y\Big)^{2/3}\\  
& + & \frac{2}{\sqrt[3]{3}}  \xi^{3/2} \frac{x^{1/6} (1+ x^2 - x^4)} {\sqrt{P} \Big(\frac{4}{3} \xi^{3/2}\sqrt{\frac{x}{P}}-y\Big)^{1/3}}\Big],
\end{eqnarray}
where $y \equiv y(x)$ and $P \equiv P(x) \equiv x^4 -3x +3$.

On the other hand, substitution of Eqs. (\ref{M-normalitzada}) and (\ref{psi-normalitzada}) in Eq. (\ref{rhok}) gives the  equation of  the $k$-lines for the $\xi$-metric:
\begin{equation}\label{k-line-normalit}
\frac{\tau_k(x)}{m} = \frac{4}{3} \xi^{3/2} \sqrt{\frac{x}{P}}- \frac{k}{3}\, \sqrt{2k} \,x^2 P.
\end{equation}
By considering appropriate initial conditions, the integration of Eq. (\ref{eq-geo}) with the second member given by  Eq. (\ref{second-member})
will be carried out with Mathematica.  The figures of this section show,  in a $(x, y)$ diagram, the resulting null geodesics coloured in red,  and also 
some representative $k$-lines: the time singularity ($k=0$) in black, the apparent horizon ($k=2$) in green, and the $k=2+\sqrt{3}$ line in blue.

 %%%%%%%%%%%%%
%           %
%  Subsection 5A  %
%           %
%%%%%%%%%%%%%

\subsection{Sufficient condition $\xi \geq 10.33$ for global nakedness}
\label{subsec-5a}

For the particular value of $\xi$,  $\xi =10.33$, some of these geodesics have been drawn in Fig. \ref{cas-xi=10}, using as mentioned above Mathematica. Specifically, we have considered four of them, corresponding to the initial conditions  $(1, 30)$, $(1, 34)$, $(1, 38)$, and $(1, 42)$. Then, in accordance with what has been mentioned at the end of Sec. \ref{sec:2},  there would be an actual or virtual pencil of outgoing radial null geodesics emanating from the central singularity and escaping outside the star to infinity since,  in accordance with Eq. (\ref{Sch-taulambda}), the corresponding geodesic times $\tau_g(\lambda)$ are lower than the horizon time $\tau_h(\lambda)$ (see Fig. \ref{cas-xi=10}). It is to be noticed that, according to Mathematica, in the overlapping region $x \lesssim 0.1$, the geodesic lines have been actually calculated (without extrapolation) up to at least $x \approx 10^{-4}.$ 

From this figure, looking at the kind of intersection with the null geodesics (updown or the opposite way), it is easily concluded that, for the considered $\xi$-value  ($\xi =10.33$), the apparent horizon is spacelike and that the $k=2 +\sqrt{3}$ line is timelike, in accordance with the results obtained in  Appendix \ref{ap-B}.
\begin{figure}
\centerline{
\parbox[c]{0.6\textwidth}{\includegraphics[width=0.5\textwidth]{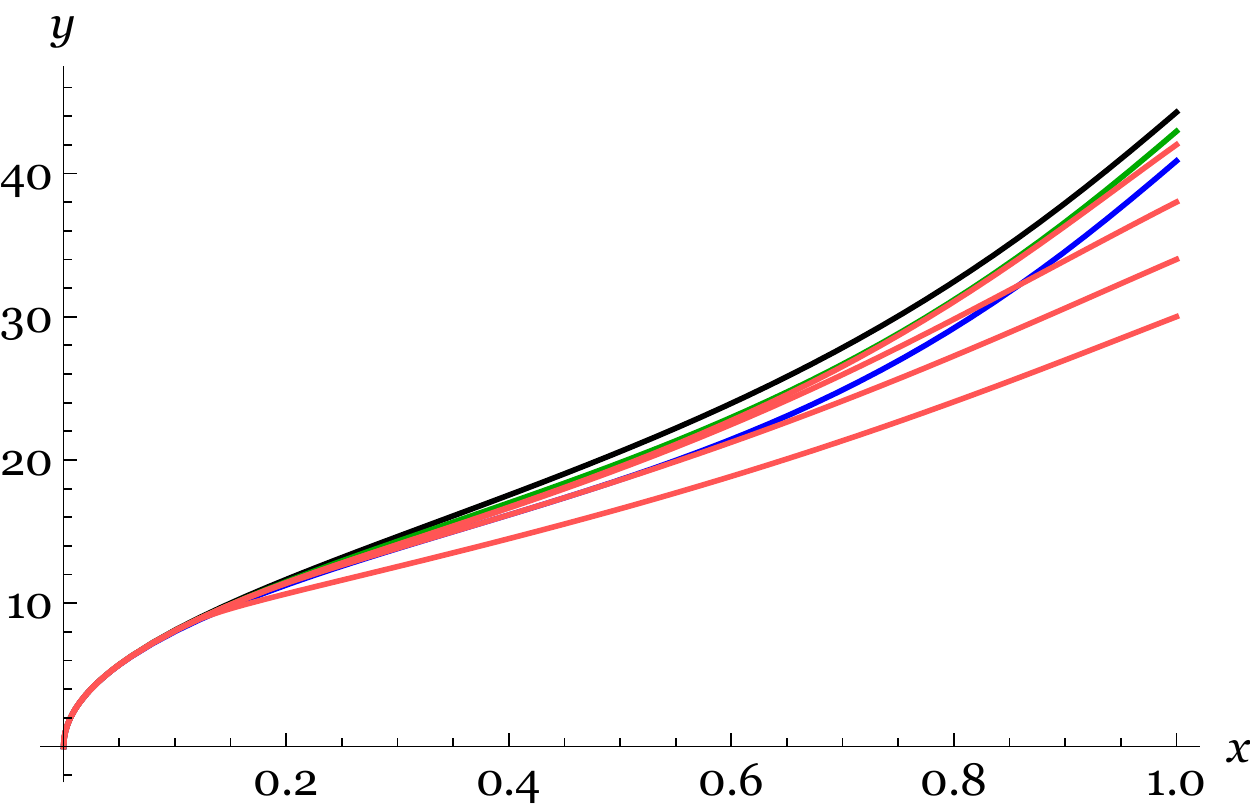}}} 
\caption{For $\xi =10.33$ the central singularity is globally naked.  The red lines are outgoing null geodesics, $(x, y(x))$, solutions of Eq. (\ref{eq-geo}) with $\xi = 10.33$. All of them would come from the central singularity: up to $x \approx 10^{-4}$ the four geodesics have actually been calculated without extrapolation. The other three lines, upper, middle and lower curves, are the singularity time (in black), the apparent horizon (in green), and the $k=2+\sqrt{3}$ line (in blue), respectively. Normalized variables, $x\equiv\rho/\lambda$ and $y\equiv \tau/m$ are used (in this and the remaining figures).\label{cas-xi=10}}\end{figure}
%

%%%%%%%%%%%%%
%           %
%  Subsection 5B  %
%           %
%%%%%%%%%%%%%

\subsection{Threshold value $\xi \approx 4.5$ for global nakedness}
\label{subsec-5b}

In a similar way, helped by Mathematica, we can find the $\xi$ values for which the central singularity becomes dressed, that is, nonglobally naked. For the particular value of $\xi$,  $\xi = 1$, one has $\tau_h(\lambda) =0$ and every null geodesic, if any, starting from the center $\rho = 0$ at $\tau = 0$ cannot reach the exterior region of the star, and then  the central singularity is, indeed, dressed (see Fig. \ref{cas-xi=1}).  Notice how the radial null geodesics leaving 
$x=0$ before $y=0$ finish their run in the intrinsic singularity time, such that the sooner the initial value of $y$ gets close to zero, the faster the geodesic runs into the 
singularity time. From Fig. \ref{cas-xi=1}, the kind of intersection of these geodesics with the apparent horizon line makes evident, in this case, that this line is spacelike.
\begin{figure}
\centerline{
\parbox[c]{0.6\textwidth}{\includegraphics[width=0.5\textwidth]{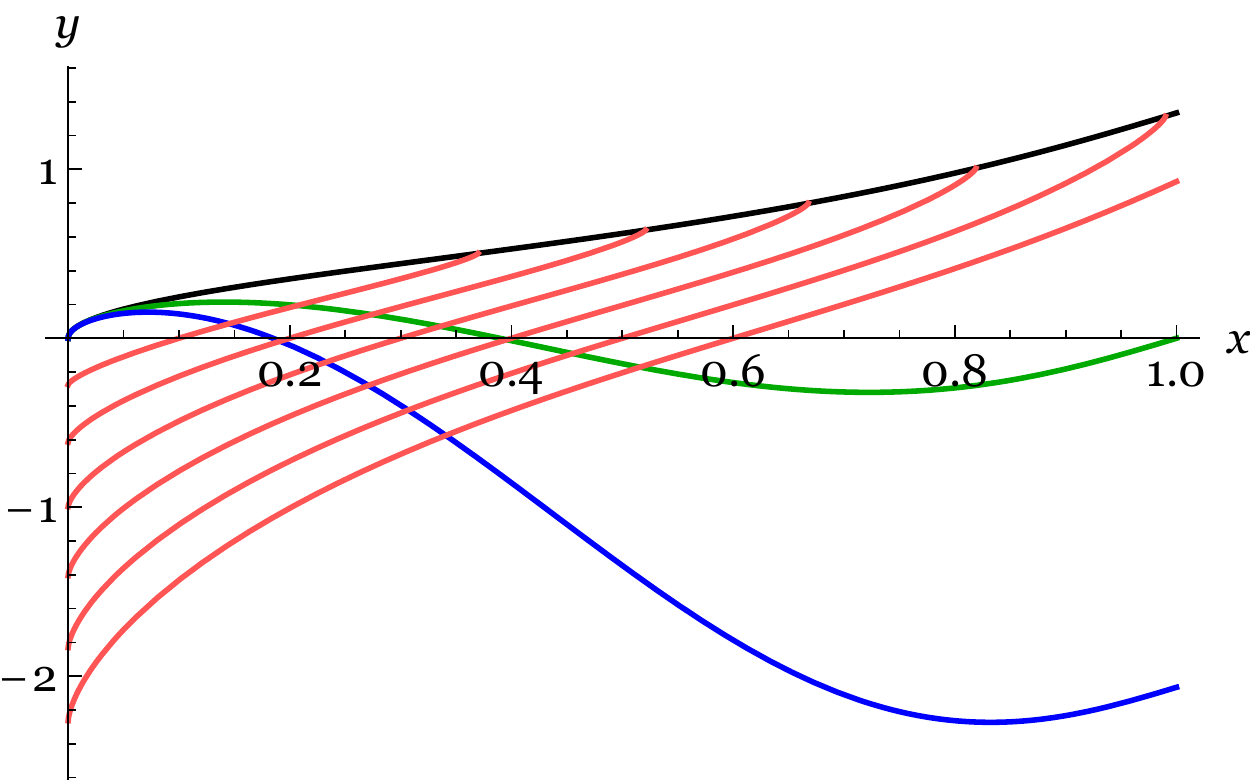}}} 
\caption{For $\xi =1$ the central singularity is dressed (nonglobally naked). The black (respectively green, blue) line stands for the singularity time (resp. apparent horizon, $k=2+\sqrt{3}$) line. The red lines are outgoing null geodesics, $(x, y(x))$, solutions of Eq. (\ref{eq-geo}) with $\xi = 1$, and correspond to the initial conditions 
$(0.1, 0), (0.2, 0), (0.3, 0), (0.4, 0), (0.5, 0), and (0.6, 0)$, respectively.  \label{cas-xi=1}}\end{figure}

The same conclusion follows by taking $\xi = 2, 3, 4$. Some outgoing radial null geodesics are plotted in Fig. \ref{cas-xi=4} for $\xi =4$, from which one sees that, when these null geodesics $\tau_g(\rho)$ approach more and more the one leaving out $\rho = 0$ at $\tau = 0$, their corresponding $\tau_g(\lambda)$ values approach, from the low, the $\tau_h(\lambda)$ value until over a certain degree of this approaching $\tau_g(\lambda)$ becomes larger than $\tau_h(\lambda)$. As a result, the corresponding photons leaving the central singularity cannot reach the exterior of the star. Further, for all the above cases with $\xi < 4$, we obtain that the central singularity is dressed too. 

\begin{figure}
\centerline{
\parbox[c]{0.6\textwidth}{\includegraphics[width=0.5\textwidth]{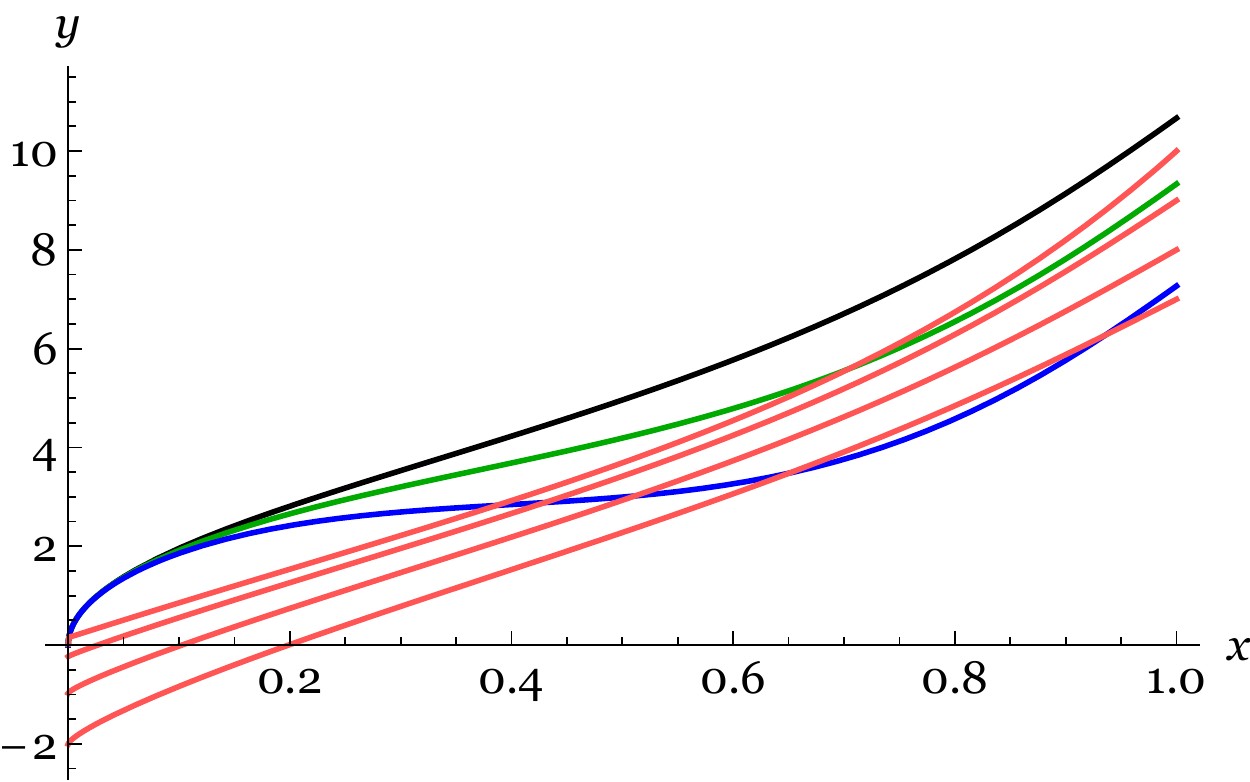}}} 
\caption{For $\xi =4$ the central singularity is dressed (non globally naked). The black (resp. green, blue) line stands for the singularity time (resp. apparent horizon, $k=2+\sqrt{3}$) line. The red lines are outgoing null geodesics, $(x, y(x))$, solutions of Eq. (\ref{eq-geo}) with $\xi = 4$, and correspond to the initial conditions  $(1, 10), (1, 9), (1, 8), and (1, 7)$, respectively. 
\label{cas-xi=4}}\end{figure}

\begin{figure}
\centerline{
\parbox[c]{0.6\textwidth}{\includegraphics[width=0.5\textwidth]{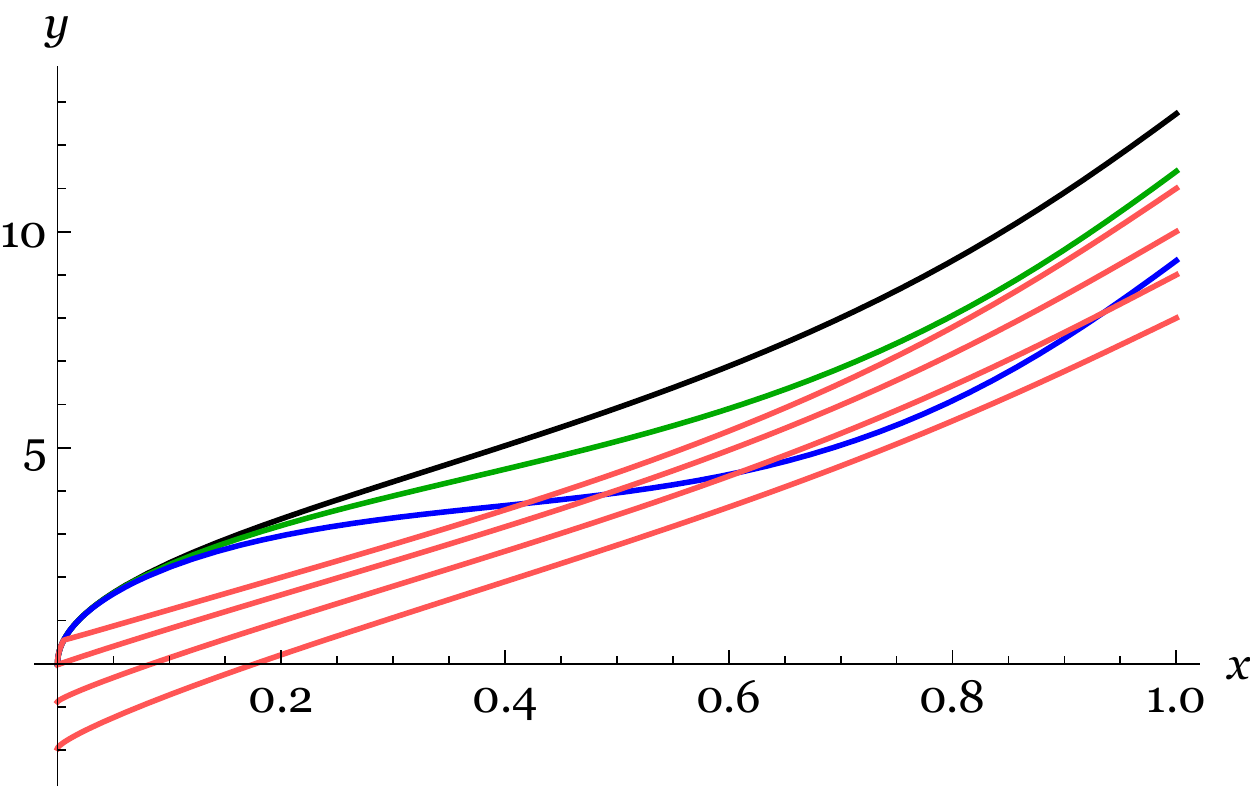}}} 
\caption{For $\xi =4.5$ the central singularity is globally naked. The black (resp. green, blue) line stands for the singularity time (resp. apparent horizon, $k=2+\sqrt{3}$) line. The red lines are outgoing null geodesics, $(x, y(x))$, solutions of Eq. (\ref{eq-geo}) with $\xi = 4.5$, and correspond to the initial conditions  $(1, 11), (1, 10), (1, 9), and (1,8)$, respectively. 
\label{cas-xi=4.5}}\end{figure}

Nevertheless, going ahead the numerical integration of Eq. (\ref{eq-geo}) with Eq. (\ref{second-member}), one can see that for $\xi = 4.5$  and $\xi =5, 6, ...$ the central singularity becomes globally naked. For $\xi =4.5$ (see Fig \ref{cas-xi=4.5}) four representative geodesics are displayed after numerical integration of Eq. (\ref{eq-geo}) considering the initial conditions $(1, 11)$, $(1, 10)$, $(1,9)$, and (1,8). The upper geodesic is the one that corresponds to the initial condition $(1, 11)$. This geodesic comes from the central singularity and escapes out of the star. Then, for this $\xi=4.5$ value, the central singularity becomes globally naked. 

A management by trial and error of the cases $4 < \xi <5$ leads to the the following result:

{\em For the $\xi$-metric,  there is a threshold $\xi$ value, say, $\xi_{0} \approx 4.5$, 
from which the central singularity becomes globally naked.}

The $\xi$-parameter, $\xi =\lambda /2m$, is related to the proper time, $\psi(\lambda)$, at which the collapsing star surface reaches the essential singularity. This follows from Eq. (\ref{psi}) by taking $\rho = \lambda$ and $M(\lambda) = m$,
\begin{equation}\label{psi-lambda}
\psi(\lambda) = \frac{2}{3} \frac{\lambda^{3/2}}{\sqrt{2m}} = \frac{4}{3} \,  m \, \xi^{3/2}.
\end{equation}
Then, we have
\begin{equation}\label{psi-lambda}
\xi = \Big(\frac{3 \tau_\lambda}{4m}\Big)^{2/3}, 
\end{equation}
where $\tau_\lambda = \psi (\lambda)$ is the proper time duration of the collapse from the formation, $\tau=0$, of the central singularity.

To end this section, we would remark that our analysis has been concentrated on the behavior of outgoing {\em radial} null geodesics. 
The reason for this self-limitation is that, for marginally bound collapse, a singularity is censored if it is radially censored 
(see Ref. \cite{Mena-Nolan}, Proposition 8). On the other hand,  we have just constructed a $\xi$-metric model with a set of natural physical 
and mathematical requirements, and we have checked numerically the global naked character of the central singularity of the model. 
For an analytical and rigorous treatment of the behavior of the radial null geodesics in the vicinity of the singular point attached to a 
marginally bound dust collapse scenario, see Ref. \cite{Giambo-Magli}.

%
%%%%%%%%%%%%%
%           %
%  SECTION  6  %
%           %
%%%%%%%%%%%%%

\section{Final considerations}
\label{sec:6}

In accordance with our result of the previous section showing that our $\xi$-metric has a global naked  central singularity, the metrics used 
in  Refs. \cite{Eardley-Smarr-1979},  \cite{Singh-Joshi-96}, and \cite{Jhin-Kau-2014} for the dust spherical collapsing case have global central naked singularities too. But these other metrics, except one,  do not satisfy all the ${\cal C}^1$ matching requirements (actually, they do not satisfy the condition $[M''] = 0$ of Eq. (\ref{Lichne})), whereas the metric with $M$ given in Ref. \cite{Jhin-Kau-2014}, Eq. (47), and our $\xi$-metric do satisfy it. Thus, a certain nonmatching  character of those metrics in Ref. \cite{Jhin-Kau-2014} could not be the reason why they violate the Penrose conjecture since two other (${\cal C}^1$ class) metrics, the $\xi$-metric plus the one associated to $M$ given by Eq. (47) in Ref. \cite{Jhin-Kau-2014}, do violate the conjecture.

Incidentally, instead of Eq. (\ref{Mrho-simple}) we could have chosen any mass function $M(\rho)$ of the large family 
\begin{equation} \label{Mrho2} 
M(\rho) = 
\begin{cases} 
\displaystyle{m +  \sum_{k=3}^\infty M_k (1- \frac{\rho}{\lambda})^k, \quad \rho \leq  \lambda} \\
m, \quad \rho \geq \lambda 
\end{cases}
\end{equation}
with the sole restriction on the constant coefficients $M_k$ that $\sum_{k=3}^\infty M_k (1- \frac{\rho}{\lambda})^k$ converges for any 
$\rho \leq \lambda$.  Actually, any of these $M(\rho)$ functions satisfies all the requirements (\ref{Lichne}). Furthermore, these coefficients should guarantee the physical condition  $M > 0$, $M' \geq 0$, $\forall \rho >0$, and even more that $M \sim \rho^n$, $n \geq 2$, for $\rho \ll \lambda$, in order that the {\em intrinsic} energy of the corresponding metric vanishes, in accordance with what is explained in Appendix \ref{ap-A} for the $\xi$-metric. 

Future work could confirm that Eq. (\ref{Mrho2}),  with the supplementary conditions for the $M_k$ coefficients, leads to marginally bound collapsing LTB metrics with their central singularities being globally naked for some $\xi$ parameter values. For the time being, we have proven easily this statement for the interesting particular case of the $\xi$-metric. 

Further, we remark that the present paper's calculations have been performed in the particular gauge $A(0, \rho) = \rho$ (see Sec. \ref{sec:2}) largely used in the literature. However, our main result --that a null geodesic, or a pencil of null geodesics, leaving the central  singularity  of the $\xi$-metric escape to the future null infinity-- is a covariant one, and therefore gauge independent. The same can be said of similar results in the above-cited references.

Finally, there is a line of thinking, that can  be traced back to  Penrose \cite{Penrose-1999}, according to which the naked singularities found in the spherically symmetric dust case, like the ones found  in the present paper, would be mere artifacts due to the oversimplified case considered. However, this objection could not be kept since the present literature on the subject shows many cases in which naked singularities persist when pressure is added to the initial dust case, and the same literature shows another cases of this persistence when the spherical symmetry is perturbed (see, for instance, Refs. \cite{MuYoSe-1974,Torres-2012,OrPi-1987} concerning the first cases and Refs. \cite{Joshi-Krolak-1996,IgNaHa-1998}, concerning the second ones).
%
%
%%%%%%%%%%%%%%%%%%
%           %
%    ACKNOWLEDGMENTS   %%
%           %
%%%%%%%%%%%%%%%%%%

\begin{acknowledgments}

This work was supported by the Spanish "Ministerio de Econom\'{\i}a y Competitividad" and the "Fondo Europeo de Desarrollo Regional" MINECO-FEDER Project No. FIS2015-64552-P. 

\end{acknowledgments}

%%%%%%%%%%%%%
%           %
%  APPENDIX %
%           %
%%%%%%%%%%%%%

\appendix

%%%%%%%%%%%%%%%%
%Appendix A
%%%%%%%%%%%%%%%%

\section{{\em Intrinsic} energy of the $\xi$-metric}
\label{ap-A}

Expressed as a 3-volume integral, the ADM energy \cite{ADM-1962} (see also Ref. \cite{Weinberg}), $P^0$ , becomes
\begin{equation}\label{P0}
P^0 = \frac{1}{8\pi} \int \frac{\partial}{\partial \rho_i} (\partial_j g_{ij} - \partial_i g) \, d\rho_1 \, d\rho_2 \, d\rho_3, 
\end{equation}
with $i, j = 1, 2, 3$, $g \equiv \delta^{ij} g_{ij}$, $G = c = 1$, and $\rho_i$ the rectilinear coordinates associated to $(\rho, \theta, \phi)$, $g_{ij}$ being the 3-space metric components.

According with the more general situation considered elsewhere \cite{LaMo-arXiv-2016}, in the particular case of our $\xi$-metric, $P^0$ becomes
\begin{equation}\label{P0-sim}
P^0 = \frac{1}{8\pi} \int \partial_i [(A - \rho A')^2 \,  \frac{n_i}{\rho^3}] \, d\rho_1 \, d\rho_2 \, d\rho_3, \quad n_i = \frac{\rho_i}{\rho}, 
\end{equation}
which we call here its {\em intrinsic} energy, since the metric is expressed in Gauss comoving coordinates adapted to the spherical symmetry, at rest at the spatial infinity, and we call these coordinates {\em intrinsic} coordinates \cite{Lapiedra-Morales-2013,LaMo-2014}.

Then, since the integrand in (\ref{P0-sim}) is regular enough (it is continuous everywhere, except for $\rho = 0$) we can apply the Gauss theorem to the corresponding 3-volume integral and express it as a 2-surface integral on the boundary. More specifically this boundary will be made of two 2-surfaces, $\rho = + \infty$ and $\rho = \epsilon >0$ where $\epsilon$ is a positive infinitesimal quantity. Then, we will take the limit $\epsilon \to 0$. 

So, we will have
\begin{equation} \label{P0-dos-sumands}
P^0=  P^0_{\infty} + \lim_{\epsilon \to 0} P^0_\epsilon, 
\end{equation}
with
\begin{equation} \label{P0-infty}
P^0_{\infty} = \lim_{\rho \to + \infty} \frac{1}{8 \pi} \int_{S_\rho} Q \, \cos \theta d \theta d \phi = \frac{1}{2}  \lim_{\rho \to + \infty} Q, 
\end{equation}
where the double integral is calculated on the 2-sphere of radius $\rho$, $S_\rho$, and
\begin{equation} \label{P0-epsilon}
P^0_{\epsilon} = - \frac{1}{2}  Q|_{\rho = \epsilon}, 
\end{equation}
and where
\begin{equation} \label{Q}
Q \equiv \frac{1}{\rho} (A - \rho A')^2.
\end{equation}

To calculate easily the limit (\ref{P0-infty}), notice that for $\rho > \lambda$ (and so for $\rho \to \infty$) our $\xi$-metric is the Schwarzschild metric, that is, Eq. (\ref{LTB0}) with eq. (\ref{k0}) given by Eq. (\ref{SchApsi}). Then an easy calculation gives for $\rho > \lambda$
\begin{equation} \label{Q-Sch}
Q = \displaystyle{\Big(\frac{9m}{2}\Big)^{2/3}   \frac{\tau^2}{\rho \Big(\tau - \frac{2}{3} \frac{\rho^{3/2}}{\sqrt{2m}}\Big)^{2/3}}},
\end{equation}
the limit of which for $\rho \to \infty$ and $\tau$ fixed vanishes. Notice that we cannot put there $\tau \geq \psi (\rho = \lambda)$, since for this value of $\rho$ the outer spherical shell of the star has just reached its own singularity and we no longer have a classical object ruled by General Relativity. The same is partially 
true for $\tau \geq \psi (\rho = 0)$.

In all, the contribution $P^0_\infty$ to the total $P^0$ vanishes and we are left with the other contribution $\lim_{\epsilon \to 0} P^0_\epsilon$. Let us calculate it. First, according to (\ref{k0}) and (\ref{Aprima}), we can write $Q$ as
\begin{equation} \label{Q-dev}
Q = \rho \Big(\frac{9M}{2(\tau - \psi)}\Big)^{2/3} \Big[\Big(\frac{1}{\rho} 
- \frac{1}{3}\frac{M'}{M}\Big) (\tau - \psi)+ \frac{2}{3} \psi' \Big]^2.
\end{equation}

Then, we are going to calculate $P^0$ for $\tau < \psi(0)$ since, as already mentioned, for $\tau = \psi(0)$ the inner spherical shell of the star reaches the intrinsic singularity and full General Relativity begins to be not completely valid. Thus, in order to calculate $\lim_{\rho \to 0} Q$, we only have to study how  the functions $M$, $M'$, and $\psi'$, present in Eq. (\ref{Q-dev}), behave in this limit. But from (\ref{psi}) and (\ref{Mrho-simple}) it is easy to see that for $\rho/\lambda \ll 1$ the function $M$ goes like $M \sim \rho^2$ and consequently $\psi \sim \rho^{1/2}$. This entails the vanishing of $\displaystyle{\lim_{\rho \to 0} Q}$. In all, both contributions to $P^0$, present in Eq. (\ref{P0-dos-sumands}), vanish, and then $P^0$ vanishes too, which means that $P^0$, the {\em intrinsic} energy of the $\xi$-metric, is stationary and finite, as is physically required.
%
%

%%%%%%%%%%%%%%%%
%Appendix B
%%%%%%%%%%%%%%%%

\section{Causal character of the lines $A=kM$}
\label{ap-B}

In this appendix, we consider the LTB marginal bound metric (\ref{LTB0}) and analyze the causal character of  the one-parameter family of radial lines  $(\tau_{k}(\rho), \rho, \theta = const, \phi = const.)$, implicitly given by 
\begin{equation}\label{k-familia}
A(\tau_k(\rho), \rho) = k M(\rho), 
\end{equation}
where $k$ is a positive real parameter.  The performed analysis is model independent in the sense that it applies for arbitrary positive increasing functions 
$\psi(\rho)$ and $M(\rho)$. For each $k$-line, the square $v_{k}^2 \equiv g_{\mu\nu} v_k^\mu v_k^\nu $  of the tangent vector,  $v_{k}^\mu = (\tau'_{k}(\rho), 1, 0, 0)$ (greek indices running from $0$ to $1$),  is 
\begin{equation}\label{vk-quadrat}
v_{k}^2= \Big(\frac{2}{k}-1 \Big) \psi'^{2} + \frac{2}{3} \sqrt{2k} (k+ 1) \psi' M' + \frac{k^2}{9}(1-2k) M'^{2}
\end{equation}
and can be written in the suitable form
\begin{equation}\label{vk2}
v_{k}^2 = M'^2 P_k(\beta), 
\end{equation}
with 
\begin{equation}\label{Pbeta}
P_k(\beta) = \Big(\frac{2}{k}-1 \Big) \beta^2 + \frac{2}{3} \sqrt{2k} (k + 1) \beta + \frac{k^2}{9}(1-2k),
\end{equation}
and where $\beta$ is a function of $\rho\in(0,\lambda)$ given by the ratio of the derivatives of the free functions of the metric (\ref{LTB0}), 
\begin{equation}\label{beta}
\beta \equiv \beta(\rho) \equiv \frac{\psi'}{M'}(\rho).
\end{equation}
The particular value $k=2$ corresponds to the apparent horizon line, in which case (\ref{Pbeta})
 becomes linear in $\beta$, $P_2(\beta) = 4(\beta -\frac{1}{3})$. The detailed analysis of the causal character of the apparent horizon for the $\xi$-metric  is carried out at the end of this Appendix. Thus, we will concentrate here on a generic value  $k \neq 2$ for which (\ref{Pbeta}) is a quadratic function of $\beta$, the discriminant of which $\Delta_k$ is always positive
\begin{equation}\label{Delta}
\Delta_k = 4 k^2,  
\end{equation}
saying that  $P(\beta)$ has two distinct real roots which can be written:
\begin{equation}\label{betes}
\beta_\varepsilon = \frac{k}{3}\, \frac{\varepsilon + \sqrt{2k}}{1 - \varepsilon \sqrt{\frac{2}{k}}}, \quad \varepsilon = \pm 1, 
\end{equation}
and then
\begin{equation}\label{beta-beta}
\beta_{+} - \beta_{-} = \frac{2k^2}{k -  2} .
\end{equation}
Notice that, for $\varepsilon = + 1$, the root $\beta_{+}$  is the function $f(k)$ defined in Eq. (\ref{compara-k}), $\beta_{+} = f(k)$; moreover, from Eq.(\ref{beta-beta}), $\beta_{+} - \beta_{-}$ is positive (respectively, negative) if $k>2$ (respectively, $k<2$). In addition, $\beta_{+}$ is positive (respectively,  negative) for $k>2$ (respectively, $k<2$), and it becomes $\beta_{+} \to + \infty$ when $k \to 2^{+}$. On the other hand, for $\varepsilon = - 1$,  the root $\beta_{-}$ is positive (respectively negative) for $k>1/2$ (respectively, $k<1/2$), and vanishes for $k=1/2$; it is finite for $k=2$, becoming for this value the root $\beta = 1/3$ of the linear polynomial  $P_2(\beta)$.

According with this analysis, we conclude that:\\

{\em For each  $k >2$ (respectively, $k<2$) and for each $\rho \in (0, \lambda)$, the line $\tau_{k} (\rho)$ is timelike 
if, and only if,  $\beta >  \beta_{+}$ or $\beta <  \beta_{-}$ (respectively,  $\beta_{+} < \beta < \beta_{-}$), with 
$\beta_{\pm}$ given by Eq. (\ref{betes}). This line is null for  $\beta = \beta_{+}$ or $\beta = \beta_{-}$, and it becomes spacelike when 
$\beta_{-} < \beta < \beta_{+}$ (respectively,  $\beta > \beta_{-}$ or $\beta < \beta_{+}$)}, where $\beta \equiv \beta(\rho)$ is given by Eq. (\ref{beta}).\\

Notice that the first member of Eq. (\ref{compara-k}), $\psi'/M'$, is bounded for every fixed $\rho \in(0, \lambda)$, but the second member, $\beta_{+}$, diverges when $k \to 2^{+}$. In fact, for  $k \to 2^{+}$,  $\beta < \beta_{+} = + \infty$ and $\beta_{-} \to 1/3$. Then,  if $\beta > 1/3$ for all $\rho \in (0, \lambda)$ the lines $\tau_{2+\epsilon} (\rho)$ are spacelike when $\epsilon \to 0^{+}$.

The apparent horizon line $A=2M$ has to be considered as a special case: for $k=2$,  Eq. (\ref{vk-quadrat}) reduces to
\begin{equation}\label{v-quadrat}
v_2^2 = 4 M' (\psi' - \frac{1}{3} M').
\end{equation}
Thus, for  $\rho \neq  0, \lambda$, the apparent horizon is spacelike, null or timelike if $\beta(\rho)$ is greater than, equal to,  or less than $1/3$, respectively.

Finally, we consider the $\xi$-metric. Deriving Eqs. (\ref{M-normalitzada}) and (\ref{psi-normalitzada}),  the function $\beta$ given by Eq. (\ref{beta}) becomes
\begin{equation}\label{beta-msimple}
\beta(x) = \frac{1}{3} \, \Big(\frac{\xi}{F(x)}\Big)^{3/2}, 
\end{equation}
where $F(x)$ is given by Eq. (\ref{Fx}). Then, using Eq. (\ref{beta-msimple}) one can express (for each $k$ value) the above results about the causal character of lines $A=kM$ in terms of the normalized variable $x = \rho/\lambda$ and the $\xi$ parameter values. In particular, the following 
statements directly result from the previous analysis.

{\em For the $\xi$-metric}:
\begin{enumerate}
\item[(i)] {\em The line $A = k_{m} M$, $k_{m} = 2 + \sqrt{3}$,  is timelike  for all $x \in (0, 1)$ if $\xi > (2 + \sqrt{3})^2 F_{max} \approx 10.33$.}
\item[(ii)] {\em The apparent horizon is spacelike for all $x \in (0, 1)$ if, and only if, $\xi > F_{max} \approx 0.74$.}
\end{enumerate}

Moreover,  for  $\xi = F_{max}$  the apparent horizon line is null at the sole point $x = x_0 \approx 0.4 \in (0,1)$, such that $F(x_0) = F_{max}$, being  spacelike $\forall x \in (0, x_0)\cup (x_0, 1)$. Otherwise,  for each $\xi < F_{max}$ there always exist two  different values, say $x_1$ and $x_2$, such that $x_1 < x_0 < x_2$ where the apparent horizon is null; of course, it is spacelike $\forall x \in (0, x_1)\cup (x_2, 1)$ and timelike $\forall x \in (x_1, x_2)$. It is tacitly understood that at $x =0$ and $x=1$ the apparent horizon line is always null whatever the value of the parameter $\xi$ may be.

%
%
%%%%%%%%%%%%%%%%%%
%           %
%    BIBLIOGRAPHY  %%
%           %
%%%%%%%%%%%%%%%%%%

\bibliography{apssamp}% Produces the bibliography via BibTeX.

\end{document}